\documentclass[aip,reprint]{revtex4-1}
\usepackage{graphicx}
\usepackage{hyperref}
\begin{document}

%\preprint{}

\title{A 17$\:$T horizontal field cryomagnet with rapid sample change designed for beamline use}


\author{Alexander T. Holmes}
\email[]{a.t.holmes@bham.ac.uk}
%\thanks{}
\author{Gary R. Walsh}
\author{Elizabeth Blackburn}
\author{Edward M. Forgan}
\affiliation{School of Physics and Astronomy, University of Birmingham, Edgbaston, Birmingham, B15 2TT, UK}
\author{Marc Savey-Bennett}
\affiliation{Cryogenic Ltd., London, UK}

\date{\today}

\begin{abstract}
We describe a new 17$\:$T cryomagnet for neutron, x-ray or optical experiments with rapid in-situ sample change. Sample temperatures are controllable from $<2\:$K to 300$\:$K in vacuum. Alternatively a room temperature bore insert can be used for experiments in the field centre under atmospheric conditions. Some of the advantages of this system include very low background scattering due to the small amount of material in the beam path, rapid cooldown, and fast field ramping. Access is available  in a $\pm 10-11^\circ$ cone around the field direction on both sides.
\end{abstract}

%\keywords{}

\maketitle

\section{Introduction}

Cryomagnets are common sample environment equipment on neutron and synchrotron beamlines.  However, there is an inherent tension between the demands of on-line scattering experiments (beam access over large solid angles, rapid sample changing and the flexibility to cope with multiple sample configurations), and those of producing high magnetic fields (small sample volume, structural reinforcement, cryogenic vacuum, cooling for magnet and sample), particularly if very uniform fields are required.

Because of this, the most common design for beamline cryomagnets has been to use split-pair coils to increase angular access.  Most typically, this has taken the form of vertical field split coils, in a pseudo-Helmholtz configuration. This permits vertical sample loading along the coil axis, and, for neutrons, close to $360^\circ$  access perpendicular to the field, (through aluminum rings that keep the coils apart). For other probes such as x-rays, different window materials are required, and these windows are invariably small, producing stress concentrations which reduce the support strength and hence maximum field available.  Other options include horizontal field split pair coils, with access in cones around the directions parallel or perpendicular to the field, to maximise available scattering angles.  The latter design in particular has very large forces between coils, limiting the maximum field available (the largest such field on a beamline to date is 11$\:$T).  In both of these configurations, the field at the sample position is less than that within the coils themselves, which limit the maximum useable field.

Certain experimental setups relax some of these constraints.  Small angle neutron scattering (SANS) is a powerful technique for investigating structures on length scales in the range 1 - 1000 nm \cite{Hammouda09}.  It has been used with great success to investigate flux line lattices (FLL) in the mixed state of Type-II superconductors \cite{forgan90,furukawa11}, where the position and intensity of the diffraction spots associated with the FLL provide unique information about the superconducting state.  The flux lines form parallel to the applied field in most cases, and hence a horizontal field, approximately parallel to the neutron beam, is most suitable to observe them easily.  These lattices show novel features up to the highest previously attainable fields \cite{Bianchi08,White11}, motivating the commissioning, from Cryogenic Ltd., of a new horizontal-field magnet capable of reaching fields of 17$\:$T.

In order to reach such a field at an affordable price and compact size, a solenoid design was adopted.  Obviously, this restricts the scattering geometry to beam directions along, or close to, the field axis, but allows much higher fields to be reached.  The chosen design also has the advantage that the superconducting coils can reach maximum field at a temperature of 4.2$\:$K, without the need for a lambda plate cooling system.  The solenoid design also has a significantly lower stray field than split coil magnets of similar field strength, which has lead to the use of actively shielded magnets in some cases \cite{Gilardi08,Allenspach09}. It also provides excellent field homogeneity (1 part in $10^4$ over a 1$\:$cm$^3$ region at the field centre), which is essential to ensure that flux lines remain parallel over the region of interest. The time taken to ramp up to full field, around 40 minutes, is also an important consideration in beamline experiments where experimental time is restricted.

In addition to high fields, low temperatures are needed to reach most superconducting states, and so the magnet is fitted with a variable temperature insert (VTI) to which the sample is connected. This sample can be changed using a horizontal loading mechanism.

As this cryomagnet has been designed with SANS in mind, another popular use of magnetic fields is to align anisotropic particles \cite{Nieh01} (e.g. colloids or polymers in solution).  These particles experience a torque favoring alignment with the field.  Accordingly, the cryomagnet can also be configured in a setup permitting access to the field centre in temperature-controlled atmospheric conditions.  Bearing in mind the limits on angular access, such a magnet has a number of other potential uses which are discussed at the end of this article.

To maximise the use of this cryomagnet, it has been designed to be transportable between large facilities, and to date, has completed six surface transports within Europe without damage, and operated successfully at both the Institut Laue-Langevin \cite{ILL2010} and the SINQ Facility at the Paul Scherrer Institut.

\section{Configuration for SANS at Low Temperature\label{LowT}}
\subsection{Magnet interior}
\begin{figure}
  \includegraphics[width=0.7\columnwidth]{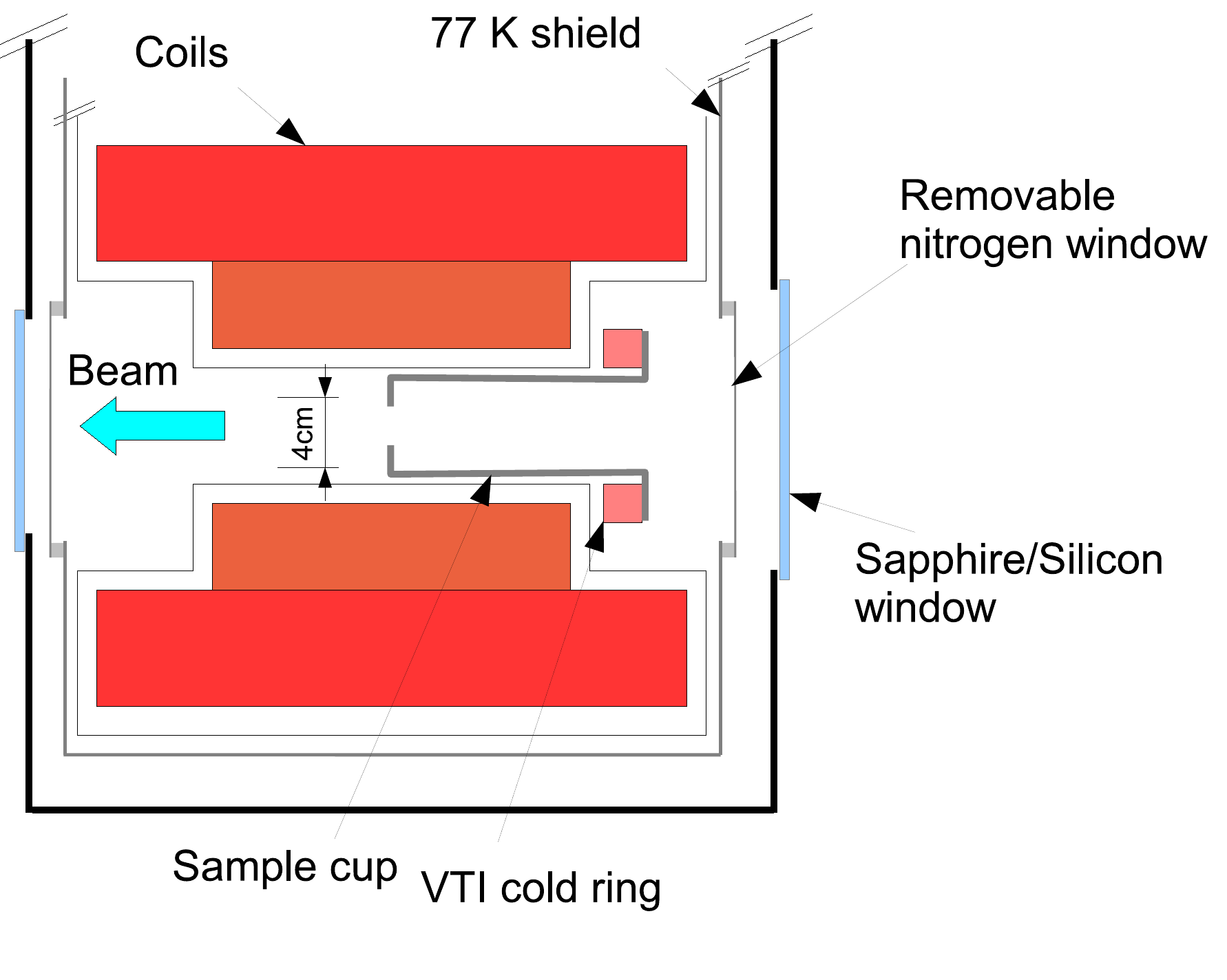}\\
  \caption{(Color online) Schematic side view cross section of the magnet bore. Not shown are the helium and nitrogen reservoirs, and an additional 40$\:$K shield cooled by helium boiloff. The windows are contained in a hinged frame, and can be sapphire or silicon, or potentially beryllium or Mylar for X-ray use.}\label{Magnet}
\end{figure}

Figure 1 shows a schematic of the interior of the magnet.  There is a cone free of obstructions on both sides of the sample position with opening angle of $\pm10^\circ$ on the upstream side, and $\pm11^\circ$ downstream, from a 10$\:$mm diameter sample.  Windows of either sapphire or silicon are currently available. Sapphire allows optical access, while silicon has a slightly lower neutron scattering background.  There are two removable radiation shields thermally connected to the nitrogen reservoir (these are referred to as `nitrogen windows' from here on).  They each consist of stainless steel and aluminum support rings holding two parallel layers of aluminum foil.  This ensures good radiation shielding whilst minimising the amount of material in the beam.

To ensure that the radiation shields and sample position remain aligned with the magnet axis as the system is cooled, the 77$\:$K shields internal to the magnet bore and the VTI ring are mechanically connected to the 4.2$\:$K magnet casing. Low thermal conductivity materials, and a long heat path are used to minimise helium boiloff.

\subsection {Variable temperature insert}
\begin{figure}
  \includegraphics[width=0.7\columnwidth]{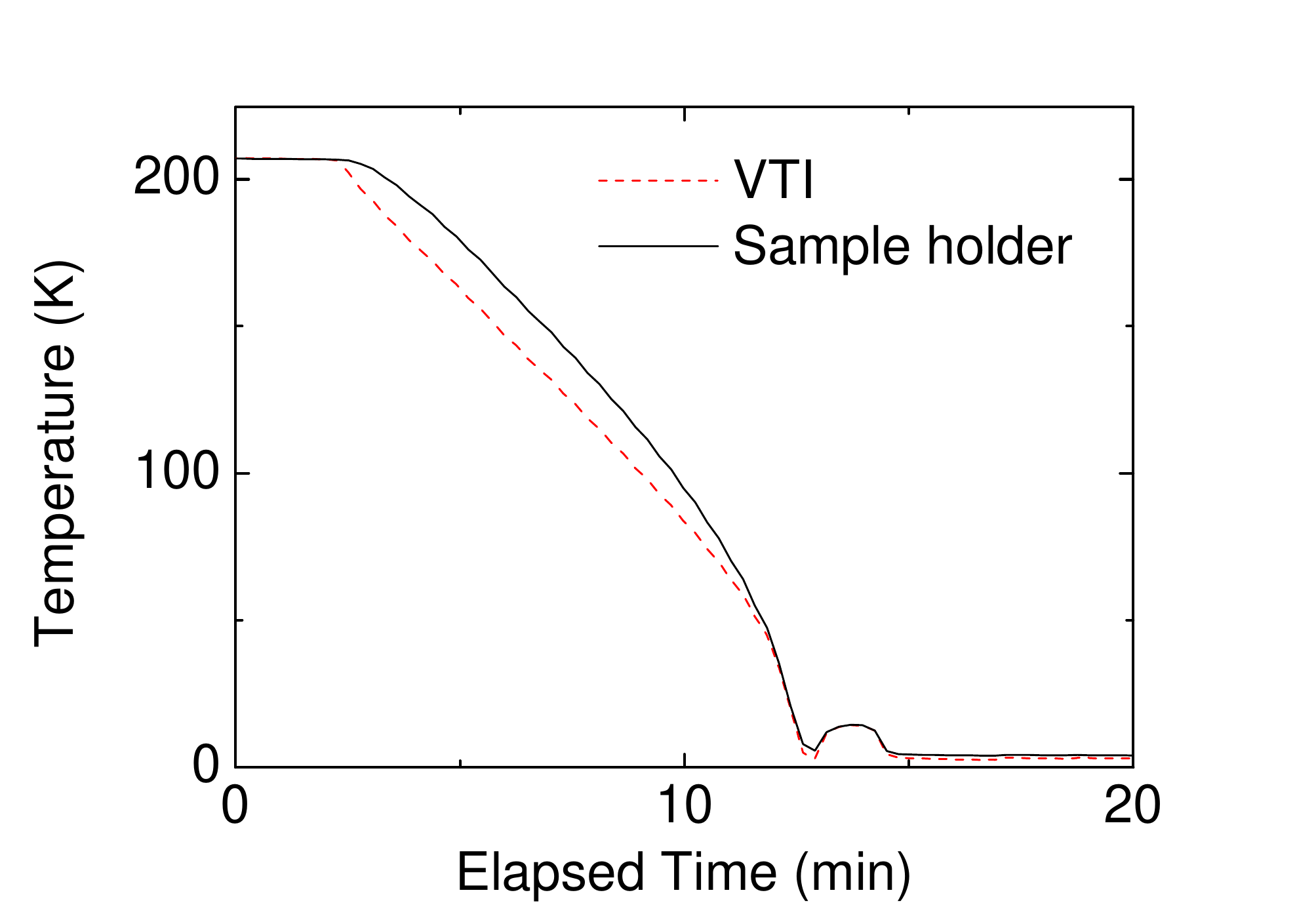}\\
  \caption{(Color online) Temperature history showing rapid cooldown of VTI and sample stage from above 200$\:$K, and fast response at low temperature. The sample holder is connected to the VTI by $\sim 10\:$cm.} \label{temphistory}
\end{figure}

Temperature control is achieved via a thermally isolated gold-plated copper ring, cooled by a brass tube fed from the helium bath, via a computer controlled needle valve. The helium is evaporated and pumped away by a rotary pump. A Cernox 1050 thermometer is positioned on the VTI cold ring, with resistive heating wrapped non-inductively around the outside of the VTI.  The temperature range of the sample stage is $\sim2\:$K up to 300$\:$K. Figure~\ref{temphistory} shows a typical cooldown and control at low temperature, demonstrating the short time to base temperature and fast thermal response.

\subsection{Sample holder}
The sample holder is a cup made of 99.99\% pure aluminum, to ensure negligible thermal gradients arising from thermal radiation from the shields. The sides contain slits to prevent significant forces arising due to eddy currents in the event of a quench. The thermal connection to the VTI cold ring is made via metal to metal contact, with three screws which can be tightened using a manipulator system as described below.  There are 16 electrical contacts available, made using push-pin style connectors which connect to contact pads on the VTI cold ring, for thermometry or other electrical connections to the sample cup. The sample itself is typically mounted on a pure aluminium plate, which is screwed on to the back of the cup facing the beam. One side of the plate may be electrically isolated to avoid creating addition eddy current paths.

\subsection{Sample Rotation mechanism}
\begin{figure}
  \includegraphics[width=0.7\columnwidth]{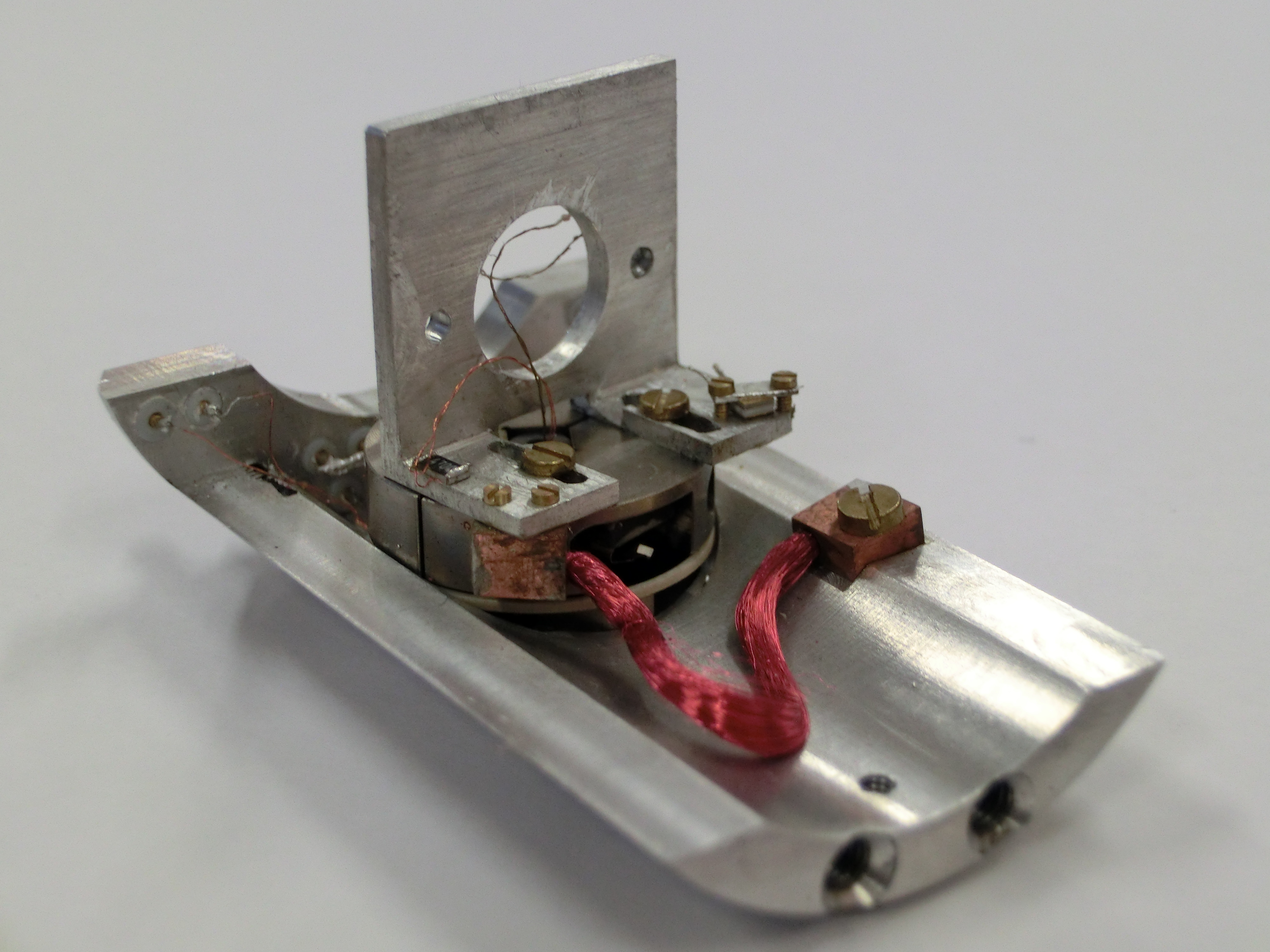}\\
  \caption{(Color online) Rotation stage holder showing attocube\textsuperscript{\textregistered}\ with 99.99\% Al sample holder, heater, thermometer, and thermal link in foreground.}\label{fig:attocube}
\end{figure}

In addition, we have a sample cup which is adapted to contain an attocube\textsuperscript{\textregistered}\ piezoelectric-driven rotation stage (Fig.~ref{fig:attocube}), which allows the sample to be rotated in situ about a vertical axis, perpendicular to the magnetic field. A potentiometer readout gives a measurement of the angle to within 0.05$^\circ$. The rotation stage is mounted on a removable platform, with push pin electrical connections onto contact pads at the back of the sample cup.  The sample itself is fixed to the top of the stage on an L-shaped holder, so that it can be turned relative to the neutron beam. Due to the intrinsically poor thermal contact from the cup to the sample holder and relatively large thermal mass, a set of 1000 25$\:\mu$m diameter copper wires was used to form a flexible thermal link, and an additional thermometer and heater were attached to the sample holder.

\section{Sample change mechanism}
\subsection{Manipulator arm}
The sample sits in the main cryostat vacuum space, and a special mechanism has been designed to change it without breaking the vacuum (and hence warming the magnet up to room temperature), which would take several days to return to a useable state. This is accomplished with a manipulator arm system, with its own vacuum chamber which fits on either end of the magnet and allows sample change to be performed under high vacuum conditions.  Before opening the access window and carrying out the sample cup removal procedure, a turbo pump or equivalent is used to reach the required vacuum level, $\sim 10^{-4}\:$mbar in the manipulator chamber; when the access window is opened with the magnet cold, cryopumping takes over and the pressure drops to $\sim 10^{-6}\:$mbar.

\begin{figure}
  \includegraphics[width=0.7\columnwidth]{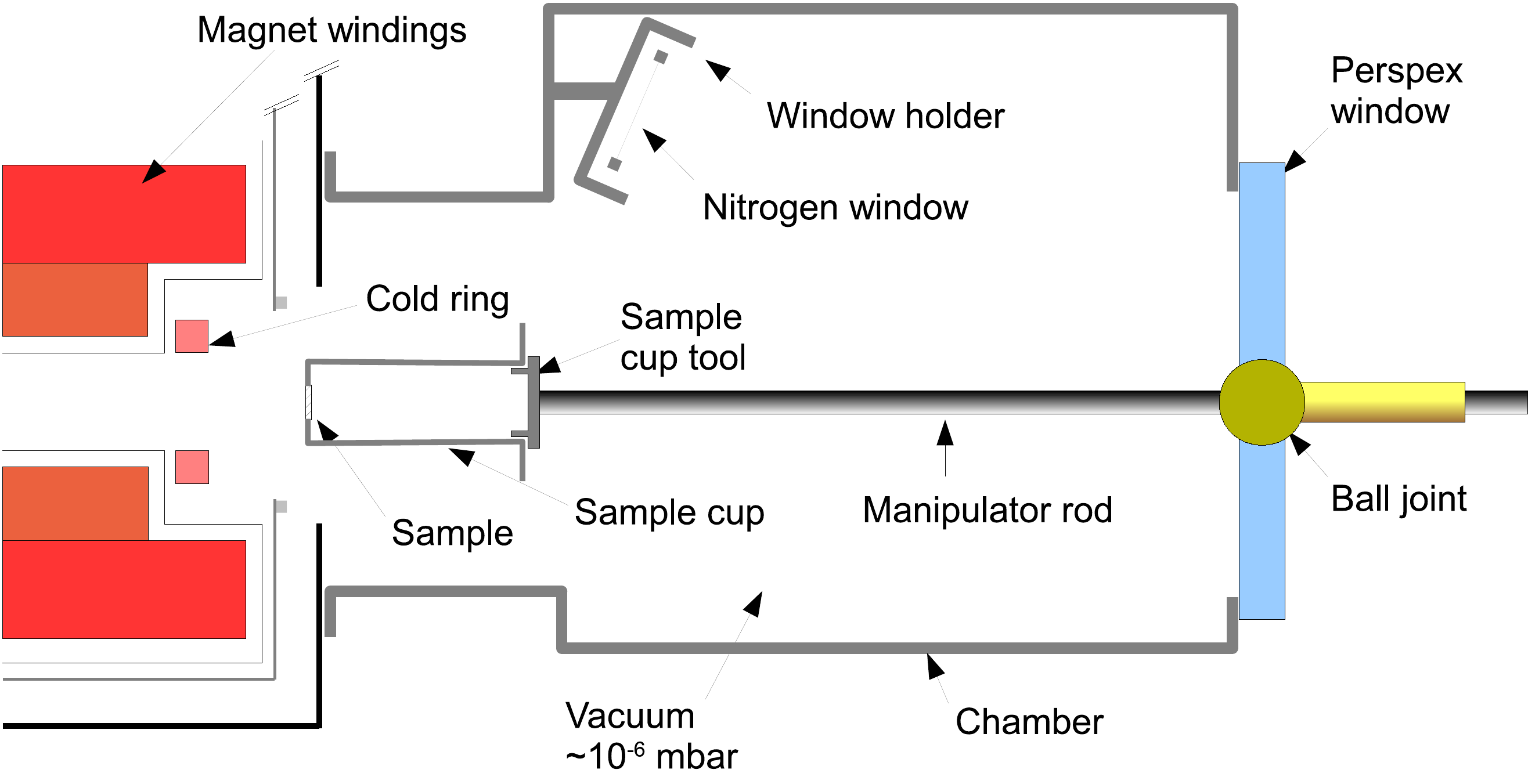}\\
  \caption{(Color online) Schematic side view of the manipulator arm and vacuum chamber. The sample cup is on the end of the manipulator arm, ready to be inserted into the VTI. The nitrogen window is shown parked in a holder. Other tools, the access window and the parking place for the sample cup tool are not shown, but are located around the inside of the manipulator chamber. O-ring seals are not shown, those in the arm are described in the text. (Not to scale.)}\label{manipulator}
\end{figure}

Figure \ref{manipulator} shows a schematic setup. The manipulator arm passes through a ball joint at the centre of a 34$\:$cm diameter, 2$\:$cm thick perspex (plexiglass) window, which allows generous visual access. The joint has four degrees of freedom; the arm can be rotated and pushed in and out as well as tilted horizontally and vertically. Double O-rings are used in the ball joint and within the arm. The space between the O-rings is pumped with a rotary pump, or connected initially to the main chamber vacuum, to give a static vacuum that minimizes leaks from air into the principal vacuum space.

On the vacuum end of the arm is a general purpose bayonet to pick up a variety of tools parked around the manipulator chamber.  The bayonet has a spring loaded locking mechanism so that tools cannot be dropped accidentally and are unlocked from the manipulator by returning them to the tool parking place. The manipulator arm also contains a push rod controlled by a micrometer gauge on the user end. This is used to open or close tools attached to the arm.
\subsection{Sample changing procedure}
We describe below the procedure for changing samples, noting important features of the system along the way:

\begin{enumerate}
  \item Attach an aluminum cover plate to the access window frame to protect the silicon/sapphire window.
  \item Attach the manipulator chamber to the cryomagnet at the input beam end and pump it out.
  \item Set the VTI temperature to at least 100$\:$K (to reduce thermal shock).
  \item Pick up a hexagonal wrench tool and unscrew the magnet window frame (the screws are held captive by the aluminum cover plate).  The window is swung open, on hinges.
  \item Loosen the screws holding the nitrogen window in place.
  \item Pick up the nitrogen window tool.  The push rod is used to control a set of sprung levered arms that can serve either to retain the tool on its holder, or grip onto the nitrogen window frame.
  \item Pick up the nitrogen window by twisting it to allow screws on the shield to pass through keyhole shaped holes on the window. Park the window in its holder inside the manipulator chamber.
  \item Loosen the M3 sample cup retaining screws on the VTI using a hexagonal wrench. These grip onto a stainless steel ring attached to the open end of the sample cup.
  \item Pick up the sample cup tool, this too has three arms which can hold the tool onto its holder, or be used to pick up a sample cup.
  \item Pick up the sample cup, by rotating the stainless steel ring 10$^\circ$, allowing the retaining screws to pass through keyhole shaped screws in the ring. Park it in a holder in manipulator chamber.
  \item Pick up a new sample cup from a second holder, and install into the VTI, engaging the keyhole shaped holes in the locking ring with the retaining screws by turning 10$^\circ$.  Pegs on the VTI ring ensure that the electrical contacts are properly aligned, and that the cup is held while the ring rotates.
  \item Tighten the retaining screws onto the cup, ensuring enough pressure to provide good thermal contact.
  \item Replace the nitrogen window and tighten the retaining bolts.
  \item Close the access window and tighten the bolts.
  \item Break the manipulator chamber vacuum and remove it.

\end{enumerate}

The entire operation, including rotation of the cryostat on the beamline by $90^\circ$ to allow access to the beam input window, and pumping of the chamber, takes about an hour. After sample change, colling to base temperature is rapid (see Fig.~\ref{temphistory} because of good metallic contact between the VTI and sample cup. The design of the manipulator chamber allows additional tools to be added should the need arise, subject only to limitations of space. However, the attocube arrangement could be used to rotate several different samples mounted on its circumference into the beam.

Automated sample changing mechanisms such as those described in Ref.~\onlinecite{Rix07} would only be viable if the vacuum chamber were permanently attached. The horizontal access and lack of space on most small angle scattering instruments precludes this. However, the attocube arrangement could be used to rotate several different samples mounted on its circumference into the beam.

\section{Configuration for ambient temperature access to field centre\label{RoomT}}

\begin{figure}[h]
   \includegraphics[width=0.7\columnwidth]{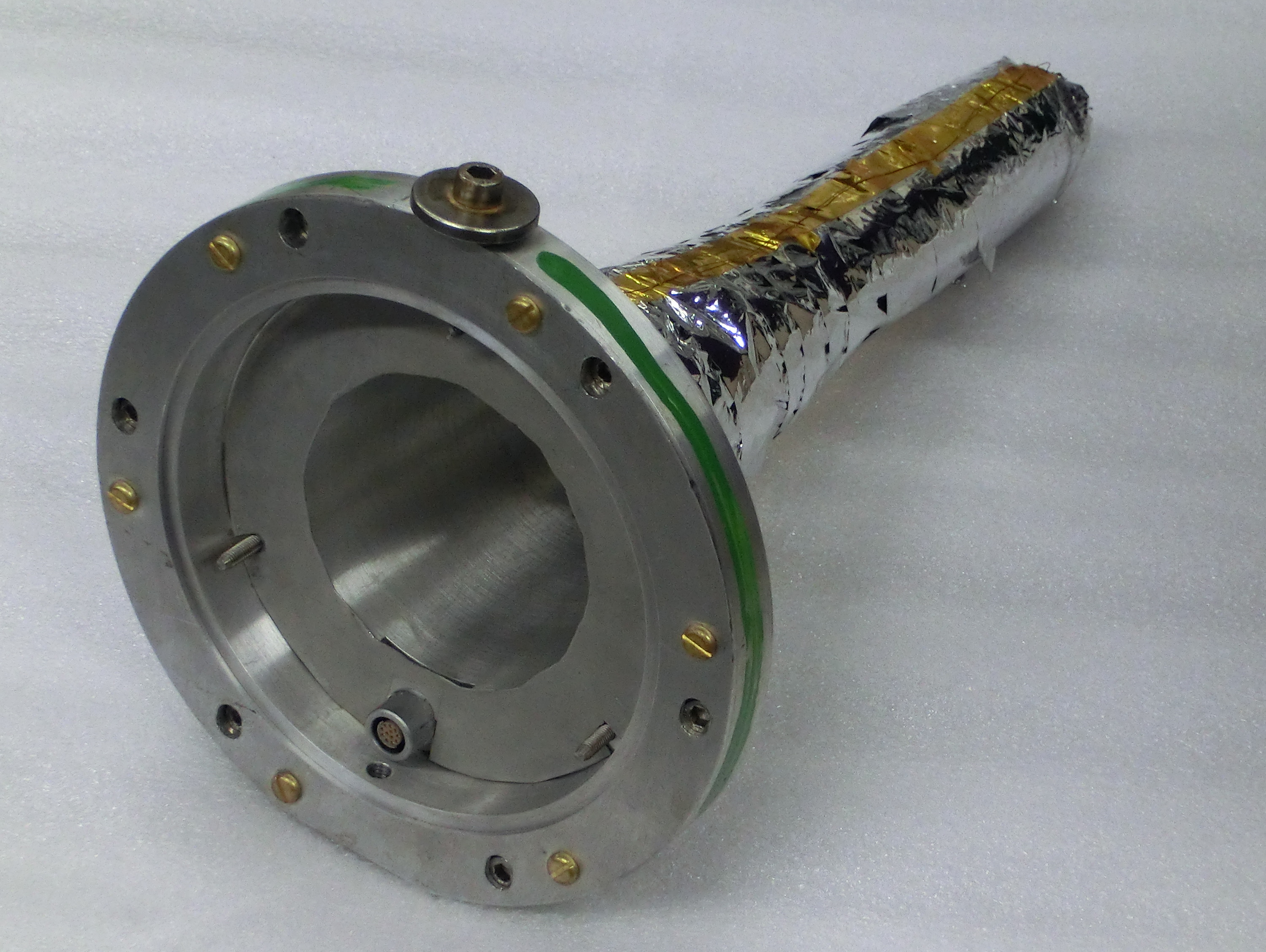}
 \caption{(Color online) Room temperature bore insert. This replaces one access window, and contains a heater and thermometer at the sample end.}
 \label{MagInsertAssembly}
\end{figure}

For experiments which require ambient pressure and temperature conditions at the field centre, one of the windows may be replaced with a conical insert made of aluminum alloy (Fig. \ref{MagInsertAssembly}).  This has a sapphire window at one end, sealed with indium.  If an unobstructed optical view along the axis is required, a sapphire window is placed in the downstream direction, and the 77$\:$K radiation shield is also removed at that end. The angular access upstream is reduced to $\pm7^\circ$ with the insert in place; the downstream angular access is unaffected.

This may be inserted into the bore, or removed, either by hand with the magnet at room temperature, or under vacuum using the nitrogen window tool. The sample is held in an aluminum puck which fits onto two pegs at the end of the bore. The sample puck can be rapidly inserted and removed with a simple insertion rod.

In order to maintain a constant temperature and prevent condensation due to radiative losses to the surrounding 77$\:$K shield, the insert has an integrated heater and platinum thermometer, as well as being surrounded with superinsulation.  If controlled cooling below room temperature is required, an aluminium tube can be fitted into the conical part of the bore, and cold nitrogen gas passed through. This cuts down the input angle, but allows rapid and accurate temperature control in the range $10-50\:^\circ$C.

To accommodate differential thermal expansion of the magnet and support system with respect to the outer vacuum vessel when the cryostat is initially cooled from room temperature, the insert can be moved vertically by 2$\:$mm without breaking the vacuum seal.

\section{Example Data}

\begin{figure}
   \includegraphics[width=0.64\columnwidth]{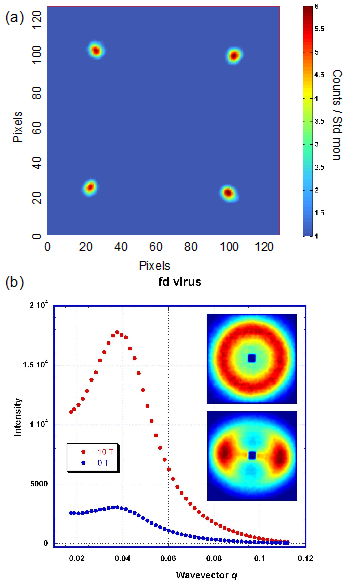}
 \caption{(Color online) (a) Diffraction pattern of the flux lattice in YBa$_2$Cu$_3$O$_7$ at 16.7$\:$T - a world record, measured at on D22, ILL. (b) Scattering versus momentum transfer for \emph{fd} virus.  The scattering from randomly oriented rods is greatly increased when they are aligned parallel to a 10 T field which is parallel to the neutron beam, giving the cylindrically symmetric pattern in the upper inset.  When the field is reduced to zero, the rods form a phase which spontaneously breaks cylindrical symmetry (lower inset). } \label{examples}
\end{figure}

Figures \ref{examples}(a) and (b) show example diffraction data taken with the sample in vacuum and using the room temperature insert, respectively. They illustrate the two main applications for this magnet --- the study of vortex lattices in superconductors, and magnetic alignment of anisotropic particles in solution at ambient conditions.

\section{Remote interfacing}

The cryomagnet system (magnet power supply, temperature control, and motorised needle valve system is controlled by a LabVIEW\texttrademark-based program provided by the manufacturer. Because the experimental procedures at large facilities are largely automated, the 17$\:$T magnet control system  has to be integrated with existing instrument control.  This is accomplished using an RS232 serial link, which allows remote control of the cryomagnet and VTI, with monitoring of all required parameters.

\section{Transportation}

The cryomagnet is specifically designed to be transportable by road between multiple facilities. It was specified to cope with an acceleration of up to $3g$ horizontally.  For protection additional to the fixed stays which hold the He bath in place, transport bolts can be inserted into the coils when at room temperature to hold it in the fixed position relative to the outer case.  The entire system fits in four reusable cases and can be transported using a small vehicle, such as a transit van.

\section{Further uses and developments.}

As well as SANS, the cryomagnet is suitable for a number of other types of experiment, particularly small angle X-ray scattering, X-ray magnetic circular dichroism (XMCD), and spin dependent Compton scattering.  Depending on the energy, this may require the replacement of the windows with Be, Al or mylar, and LN$_2$ radiation shields with aluminised mylar. We are currently in discussions to carry out SAXS measurements in the near future.

Intended future developments would be to retrofit a $^3$He or dilution refrigerator unit.  The latter would somewhat restrict the angular range, but could use the existing VTI system as the 1$\:$K pot.

At present the magnet is in use at European neutron sources, and enquiries about collaboration are welcomed.

\acknowledgments{}
The authors acknowledge the EPSRC for providing funding for this project, the Institut Laue-Langevin, and Paul Scherrer Institute for their help in integrating the magnet at their facilities, particularly the instrument responsibles Charles Dewhurst (D22), Jorge Gaviliano, Joachim Kohlbrecher (SANS-I) and Ralf Schweins (D11).  We also thank the numerous technicians, sample environment and software experts who have helped, including Marek Bartkowiak, Dave Bowyer, Ken Honniball, Eddy Leli\`{e}vre-Berna, Paulo Mutti, and Markus Zolliker.

%\bibliography{refs}
%

\end{document}